\documentclass[sigconf]{acmart}

\AtBeginDocument{%
 }

\copyrightyear{2023}
\acmYear{2023}
\setcopyright{rightsretained}
\acmConference[CIKM '23]{Proceedings of the 32nd ACM International Conference on Information and Knowledge Management}{October 21--25, 2023}{Birmingham, United Kingdom}
\acmBooktitle{Proceedings of the 32nd ACM International Conference on Information and Knowledge Management (CIKM '23), October 21--25, 2023, Birmingham, United Kingdom}\acmDOI{10.1145/3583780.3615281}
\acmISBN{979-8-4007-0124-5/23/10}

\makeatletter
\gdef\@copyrightpermission{
  \begin{minipage}{0.3\columnwidth}
   \href{https://creativecommons.org/licenses/by/4.0/}{\includegraphics[width=0.90\textwidth]{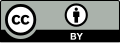}}
  \end{minipage}\hfill
  \begin{minipage}{0.7\columnwidth}
   \href{https://creativecommons.org/licenses/by/4.0/}{This work is licensed under a Creative Commons Attribution International 4.0 License.}
  \end{minipage}
  \vspace{5pt}
}
\makeatother

\usepackage{nicefrac}
\usepackage{xfrac}
\usepackage[capitalise]{cleveref}
\usepackage{pgfplotstable}
\usepackage{pgfplots}
\usepackage{tikz}
\pgfplotsset{compat=1.9}
\usepackage{subcaption}
\usepackage{xcolor}

\begin{document}
\bibliographystyle{plainnat}

\title[Test-Time Embedding Normalization for Popularity Bias Mitigation]{Test-Time Embedding Normalization \\ for Popularity Bias Mitigation}


\author{Dain Kim}
\email{dain5832@postech.ac.kr}
\orcid{1234-5678-9012}
\authornote{Both authors contributed equally to this research.}
\affiliation{%
  \institution{GSAI, POSTECH}
  \city{Pohang}
  \country{Republic of Korea}
}

\author{Jinhyeok Park}
\email{jinhyeok1234@postech.ac.kr}
\authornotemark[1]
\affiliation{%
  \institution{GSAI, POSTECH}
  \city{Pohang}
  \country{Republic of Korea}
}
\author{Dongwoo Kim}
\email{dongwoo.kim@postech.ac.kr}
\affiliation{%
  \institution{CSE \& GSAI, POSTECH}
  \city{Pohang}
  \country{Republic of Korea}
}

\renewcommand{\shortauthors}{Dain Kim, Jinhyeok Park, \& Dongwoo Kim}


\begin{abstract}
Popularity bias is a widespread problem in the field of recommender systems, where popular items tend to dominate recommendation results. In this work, we propose ‘Test Time Embedding Normalization’ as a simple yet effective strategy for mitigating popularity bias, which surpasses the performance of the previous mitigation approaches by a significant margin. Our approach utilizes the normalized item embedding during the inference stage to control the influence of embedding magnitude, which is highly correlated with item popularity. Through extensive experiments, we show that our method combined with the sampled softmax loss effectively reduces popularity bias compare to previous approaches for bias mitigation. We further investigate the relationship between user and item embeddings and find that the angular similarity between embeddings distinguishes preferable and non-preferable items regardless of their popularity. 
The analysis explains the mechanism behind the success of our approach in eliminating the impact of popularity bias. 
Our code is available at \url{https://github.com/ml-postech/TTEN}.
\end{abstract}

\ccsdesc[500]{Information systems~Recommender systems}
\keywords{recommender systems, popularity bias, fairness}




\newcommand{\model}{TTEN}
\newcommand{\dw}[1]{{\color{red}{(#1)}}}

\maketitle
\section{Introduction}\label{intro}
In recent years, recommender systems have reached successful accomplishments in providing personalized recommendations by analyzing user history. These systems are widely employed in various domains, such as e-commerce, recruitment, and online content platforms~\cite{schafer2001commerce, farber2003automated, van2013deep}. However, the presence of popularity bias is one common challenge faced by recommender systems. 
Popularity bias refers to a phenomenon where popular items are overrepresented in the recommendation results, while less popular items receive less exposure than they deserve~\cite{abdollahpouri2019popularity}. This bias arises due to the inherent nature of recommendation algorithms where the popular items take a large portion of the train data, leading them to become more dominant in the recommendation list~\cite{abdollahpouri2019unfairness}.

In this paper, we introduce a novel strategy called `Test Time Embedding Normalization (\model)' to mitigate popularity bias in recommender systems. 
We begin by showing that there exists a significant correlation between the popularity of an item and the magnitude of its embedding. We further analyze the cosine similarity between the user and item embeddings. 
When trained with a proper loss, such as sampled softmax loss~\cite{wu2022effectiveness}, we find that the cosine similarity distinguishes preferable and non-preferable items regardless of their popularity.
Interestingly, our observation indicates that the well-known models are inherently capable of disentangling popularity and preference without the need for explicit bias mitigation algorithms~\cite{liang2016causal,wei2021model,ren2022mitigating,kou2022bignn}. The commonly used inner product score function, however, strengthens the popularity bias since the function multiplies cosine similarity and magnitude of embeddings together.
Building upon this observation, we aim to control the effect of popularity by normalizing item embeddings when generating the recommendation results. We conduct extensive experiments and find the proposed approach outperforms existing state-of-the-art popularity bias mitigation strategies.

Our contribution can be summarized as follows:
\begin{itemize}
\item We propose a test-time embedding normalization method with controllable normalization strength.
\item We show that the normalization with the sampled softmax can effectively outperform the existing approaches aimed at popularity bias mitigation.
\item Our analysis shows that the magnitude of item embeddings is highly correlated with item popularity, and cosine similarity is sufficient to capture the relevance of items.
\end{itemize}

\section{Related Work}\label{related_works}

We introduce previous studies that aim to mitigate popularity bias in recommender systems. \citet{liang2016causal} employ a re-weighting strategy that assigns weights inversely proportional to the item's popularity.
\citet{zheng2021disentangling} separate user and item embeddings into interest and conformity using a negative sampling strategy based on causality.
\citet{ren2022mitigating} propose a gradient-based method to address popularity bias. They argue that popularity bias arises due to the dominance of popular items in the positive items, resulting in a significantly larger gradient of positive items than negative items. Consequently, a gradient-adjusting algorithm is introduced to mitigate the popularity bias. \citet{wu2022effectiveness} suggest sampled softmax for recommender systems. This work shows that sampled softmax maximizes discounted cumulative gain and alleviates popularity bias as it samples popular items as negative more frequently.

\section{Methods}\label{methods}

\subsection{Preliminaries}

\textit{Problem Formulation.}
Let $\mathcal{U}=\{1, \dots, U\}$ denote the set of users and $\mathcal{I}=\{1, \dots, I\}$ the set of items. 
A set of interactions between the users and the items can be represented as a binary matrix $R \in \mathbb{R}^{U \times I}$ where $R_{ui}=1$ if user $u$ interacted with an item $i$ and $R_{ui}=0$ otherwise. 
Embedding-based models in recommender systems aim to learn user embeddings $e_u \in \mathbb{R}^{d}$ and item embeddings $e_i \in \mathbb{R}^{d}$, where $d$ is a dimension of embedding~\cite{he2017neural, koren2009matrix, he2020lightgcn}. 
Once the embeddings are obtained, the inner product between user and item embeddings is used to compute the relevance score $\hat{r}_{ui} = e_u^{\top} e_i$ between user $u$ and item $i$. Top-$k$ most relevant items are then recommended for each user.

\textit{Graph Convolution Network.} A graph convolution network (GCN) framework has recently emerged as a state-of-the-art embedding-based approach in recommendation~\citep{he2020lightgcn}. The GCN leverages the user-item interaction matrix to propagate user and item information through their interaction. The embeddings are iteratively updated through graph propagation, leading to enriched representations that capture the relationship between users and items.

\textit{Loss Functions.} A proper loss function needs to be defined to learn the user and item embeddings through the GCN. We consider two popular loss functions: Bayesian Personalized Ranking (BPR)~\cite{rendle2012bpr} and Sampled Softmax (SSM)~\cite{wu2022effectiveness} losses.

The BPR is a pairwise loss that encourages the model to rank observed user-item interactions higher than unobserved interactions.
The SSM loss maximizes the probability of a positive item among sampled negative items through a softmax function. 
Let $\mathcal{N}_u$ is a set of sampled negative items for user $u$, i.e., $R_{uj} = 0 \; \forall j \in \mathcal{N}_u$.
The SSM loss for user $u$ is formulated as follows:
\begin{equation}
\mathcal{L}_{\text {ssm }}(u)=-\sum_{i:R_{ui}=1} \log \frac{\exp (\nicefrac{f(e_u, e_i)}{\tau})}{\exp (\nicefrac{f(e_u, e_i)}{\tau})+\sum_{j \in \mathcal{N}_u} \exp (\nicefrac{f(e_u, e_j)}{\tau})}.
\end{equation}
where $f(e_u, e_i)$ is a cosine similarity of two embeddings and $\tau$ is temperature.

The negative items of each user are taken from the positive items of the other users. In doing so, popular items are sampled more frequently than unpopular items, making them the hard negatives.

\subsection{Test Time Embedding Normalization}
In this section, we begin by analyzing the relationship between the popularity of items and the magnitude of their embeddings. \cref{tab:pearson correlation} presents the correlation between item popularity and embedding magnitude, demonstrating a significant correlation between these two factors. Our finding aligns with previous work by \citet{ren2022mitigating}, who also observe that popular items tend to have larger embedding magnitudes due to positive gradients acquired during model updates.

Based on our findings, we introduce a strategy called Test Time Embedding Normalization (\model), which aims to mitigate popularity bias in recommender systems. Our approach controls the impact of item embedding magnitudes ($\ell_2$ norm) during inference, given their strong correlation with item popularity.

To recommend items, an inner product between user and item embeddings is widely used as a relevance score,
i.e., $\hat{r}_{ui} = e_u^{\top} e_i$.
By decomposing the inner product of user embeddings and item embeddings as $e_u^{\top} e_i={\cos}({e}_u, {e}_i) \|{e}_u\| \|{e}_i\|$ where $\cos(\cdot)$ is a cosine similarity, we can rewrite the inner product as their cosine similarity and magnitudes. To mitigate the impact of the magnitude, which is closely tied to item popularity, we propose \model{} to compute the relevance score as
\begin{equation}
\label{eqn:p_scale}
\hat{r}_{ui} = \frac{e_u^{\top} e_i}{\|e_u\| \|e_i\|^{p}} = \cos(e_u, e_i)\|e_i\|^{(1-p)},
\end{equation}
where $p$ controls the strength of the normalization. If $p=1$, the relevance score only depends on the cosine similarity between two embeddings. If $p=0$, the relevance score follows the inner product. Through the different choices of $p$, we can control the influence of the embedding magnitude for the recommendation. Note that the magnitude of user embedding does not influence the ranking of the final recommendation list for a given user.

The normalization process is computationally efficient and can be easily integrated into existing embedding-based recommender systems, making it a practical solution for mitigating popularity bias in real-world applications.
While previous research~\cite{wu2021self, mao2021simplex} utilizes the normalization during the training phase, the potential of normalization to mitigate popularity bias during the inference stage has been underexplored. Although some works~\cite{wu2022effectiveness, chen2023adap} compared the results of normalization during the training and test time, their tests were not conducted in the context of bias mitigation, missing the importance of normalization in mitigating popularity bias.

\section{Experiments}\label{experiments}
In this section, we evaluate the performance of our proposed method, \model. Furthermore, we conduct a comprehensive analysis to investigate the behavior of our approach through a series of experiments. We address the following research questions and provide insights into the proposed approach.

\begin{itemize}
    \item RQ1: Does the proposed test time embedding normalization outperform existing strategies for mitigating popularity bias in recommender systems?
    \item RQ2: Why does test time embedding normalization eliminate popularity bias in recommender systems?
    \item RQ3: How does the scale of the normalization impact the recommendation results of unpopular and popular items?

\end{itemize}

\subsection{Experimental Setup}
\textit{Dataset.}
We utilize three publicly available datasets, Gowalla~\cite{cho2011friendship}, Yelp2018\footnote{https://www.yelp.com/dataset}, and ML-10m\footnote{https://grouplens.org/datasets/movielens/}, which are widely recognized and employed in the field of recommendation systems. For a fair comparison, we preprocessed the Gowalla and Yelp2018 datasets following the methods described in \citet{he2020lightgcn}. For the ML10m dataset, we transform the explicit ratings into implicit feedback, assigning a value of 1 if the user rated the item and 0 otherwise, following the work of \citet{wei2021model}. The detailed statistics of these datasets are provided in \cref{tab:data_stats}.

\begin{table}[t!]
    \centering
    \caption{Statistics of the datasets.}    
    \begin{tabular}{c r r r r}
    \toprule
    \text { Dataset } & \text { \#Users } & \text { \#Items } & \text { \#Interactions } & \text { Density } \\
    \midrule
    \text { Gowalla } & 29,858 & 40,981 & 1,027,370 & 0.00084 \\
    \text { Yelp2018 } & 31,668 & 38,048 & 1,561,406 & 0.00130 \\
    \text { ML10M } & 69,166 & 8,790 & 5,000,415 & 0.00823 \\
    \bottomrule
    \end{tabular}

    \label{tab:data_stats}
\end{table}

\textit{Baselines.}
We use LightGCN (LGN)~\cite{he2020lightgcn} trained with BPR and SSM loss as a baseline and backbone model for which \model{} is applied. We compare our approach with four methods, IPS~ \cite{liang2016causal}, MACR~\cite{wei2021model}, GRAD~\cite{ren2022mitigating}, and BIGNN~\cite{kou2022bignn}, designed to mitigate the popularity bias in recommender systems. We reproduce all baselines except for GRAD and BiGNN.%

\textit{Evaulation Protocols.}
Many approaches that try to mitigate popularity bias~\cite{zheng2021disentangling, wei2021model, ren2022mitigating} commonly utilize an unbiased test set to assess the performance of the proposed method. This is because the biased test set, which often follows a long-tail distribution, can yield high performance even when the model produces a biased recommendation. Therefore, we follow evaluation protocols commonly used in related studies~\cite{zheng2021disentangling, wei2021model, ren2022mitigating} to appropriately assess the impact of removing popularity bias. The \emph{unbiased test set} is constructed by randomly sampling items from a uniform distribution, ensuring that each item has an equal probability of being selected. We use the train and test set from the MACR~\cite{wei2021model} for a fair comparison with existing methods and keep 50 \% of the test set as the validation set for hyperparameter search. 
The performance of the model is evaluated through Recall@20 and NDCG@20, considering all unobserved items as a negative set.

\textit{Implementation Details.}
The dimension of user and item embeddings are both set to  64, and the embeddings are initialized using the Xavier initializer~\cite{glorot2010understanding}. The model is trained for 300 epochs, with early stopping applied after 50 epochs with patience of five. For optimization, we utilize the Adam optimizer~\cite{kingma2014adam} with a learning rate of 1e-3 and batch size of 4096. $L_2$ regularizer coefficient is set to 1e-5 for BPR loss and 1e-7 for SSM loss. Three-layered LightGCN is used for all experiments. Following the guidelines in \citet{wu2022effectiveness}, we conduct a hyperparameter search as outlined in \citet{wu2021self} to discover the optimal temperature value. The temperature of 0.1 for the Gowalla dataset, 0.12 for the Yelp2018 dataset, and 0.1 for the ML10M dataset are chosen through the search. We set the normalization strength to one, i.e., $p=1$, unless noted.

\begin{table}[t]
    \centering
    \caption{Overall performance of our method and baselines. \\ \textsuperscript{$\dagger$} and \textsuperscript{$\ast$} are taken from original paper ~\cite{ren2022mitigating} and ~\cite{kou2022bignn}, resp.}
    \label{tab:main_results}    
    \resizebox{\linewidth}{!}{
    \begin{tabular}{c c c c c c c}
    \toprule
    \multicolumn{1}{c}{ Dataset } & \multicolumn{2}{c}{ Gowalla } & \multicolumn{2}{c}{ Yelp2018 } & \multicolumn{2}{c}{ ML10M } \\
    \midrule
    Method & Recall &  NDCG & Recall &  NDCG & Recall  & NDCG \\
    \midrule
     LGN (BPR) & 0.044 &  0.027 & 0.005 &  0.006 & 0.006&  0.004 \\
     LGN (SSM) & 0.075 & 0.045 & 0.009 & 0.010 & 0.011 & 0.007\\
     IPS & 0.057  & 0.035 & 0.012 & 0.012 & 0.018 & 0.010 \\
     GRAD\textsuperscript{$\dagger$} & 0.081 & 0.053 & - & - & - & - \\
     MACR & 0.100  & 0.054 & 0.047 & 0.027 & 0.057& 0.028\\
     BiGNN\textsuperscript{$\ast$} & 0.108  & 0.059 & 0.046 & 0.027 & 0.059& 0.032\\
    \midrule
    \model{} (BPR) & 0.081 & 0.051 & 0.017 & 0.015 & 0.022 & 0.013\\
    \model{} (SSM) & \textbf{0.142} & \textbf{0.092} & \textbf{0.049} & \textbf{0.035} & \textbf{0.065} & \textbf{0.036} \\
    \bottomrule
\end{tabular}    
}
\end{table}

\pgfplotsset{compat=1.11,
    /pgfplots/ybar legend/.style={
    /pgfplots/legend image code/.code={%
       \draw[##1,/tikz/.cd,yshift=-0.25em]
        (0cm,0cm) rectangle (3pt,0.8em);},
   },
}

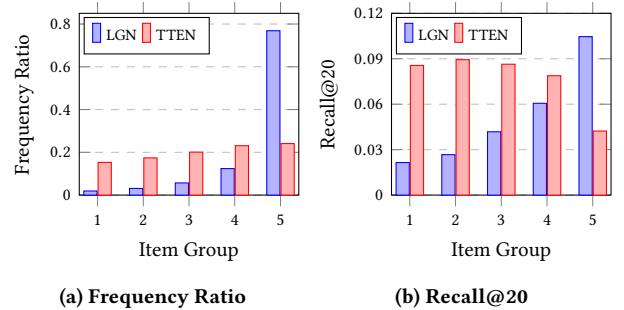
\begin{figure}[ht!]
    \centering
    \begin{subfigure}{.475\linewidth}
    \centering
\begin{tikzpicture}

\begin{axis}[
    ticklabel style = {font = {\fontsize{6 pt}{6 pt}\selectfont}},
    label style = {font = {\fontsize{8 pt}{12 pt}\selectfont}},
    legend style = {font = {\fontsize{6 pt}{12 pt}\selectfont}},
    xlabel={Item Group},
    ylabel={Frequency Ratio},
    ymin=0, ymax=0.85,
    xtick={1,2,3,4,5},
    ytick={0,0.2, 0.4, 0.6, 0.8},
    legend style={at={(0.02, 0.98)}, anchor=north west, legend columns=-1},
    ymajorgrids=true,
    grid style=dashed,
    width = 4.5cm,
    height = 4cm,
    ybar=0.4pt,
    bar width=5pt]

    \addplot
        coordinates {
        (1,0.0192)(2,0.0312)(3,0.0569)(4,0.1239)(5,0.7686)
        };
    \addplot
        coordinates {
        (1,0.153)(2,0.174)(3,0.201)(4,0.231)(5,0.241)
        };
    \legend{LGN, \model}
    \end{axis}
\end{tikzpicture}
\caption{Frequency Ratio}
\end{subfigure}
\begin{subfigure}{.475\linewidth}
    \centering
    \begin{tikzpicture}
\begin{axis}[
    ticklabel style = {font = {\fontsize{6 pt}{6 pt}\selectfont}},
    label style = {font = {\fontsize{8 pt}{12 pt}\selectfont}},
    legend style = {font = {\fontsize{6 pt}{12 pt}\selectfont}},
    xlabel={Item Group},
    ylabel={Recall@20},
    ymin=0, ymax=0.12,
    xtick={1,2,3,4,5},                  
    ytick={0,0.03, 0.06, 0.09, 0.12},
    yticklabel style={
        /pgf/number format/fixed,
        /pgf/number format/precision=5},
    legend style={at={(0.02, 0.98)}, anchor=north west, legend columns=-1},
    ymajorgrids=true,
    grid style=dashed,
    width = 4.5cm,
    height = 4cm,
    ybar=0.4pt,
    bar width=5pt]
\addplot
    coordinates {
    (1,0.0215)(2,0.0267)(3,0.0418)(4,0.0606)(5,0.1046)
    };
\addplot
    coordinates {
    (1,0.0856)(2,0.0894)(3,0.0864)(4,0.0788)(5,0.0423)
    };
\legend{LGN, \model}
\end{axis}
\end{tikzpicture}
\caption{Recall@20}
\end{subfigure}
\caption{Frequency and recall@20 of each item group in the Gowalla dataset. The SSM loss is used with LightGCN (LGN).}
\label{fig:item_group_freq}
\vspace{-5mm}
\end{figure}

\subsection{Overall Performance (RQ1)}

\begin{figure*}[ht]
     \centering
     \begin{subfigure}[b]{0.23\textwidth}
         \centering
         \includegraphics[width=\textwidth]{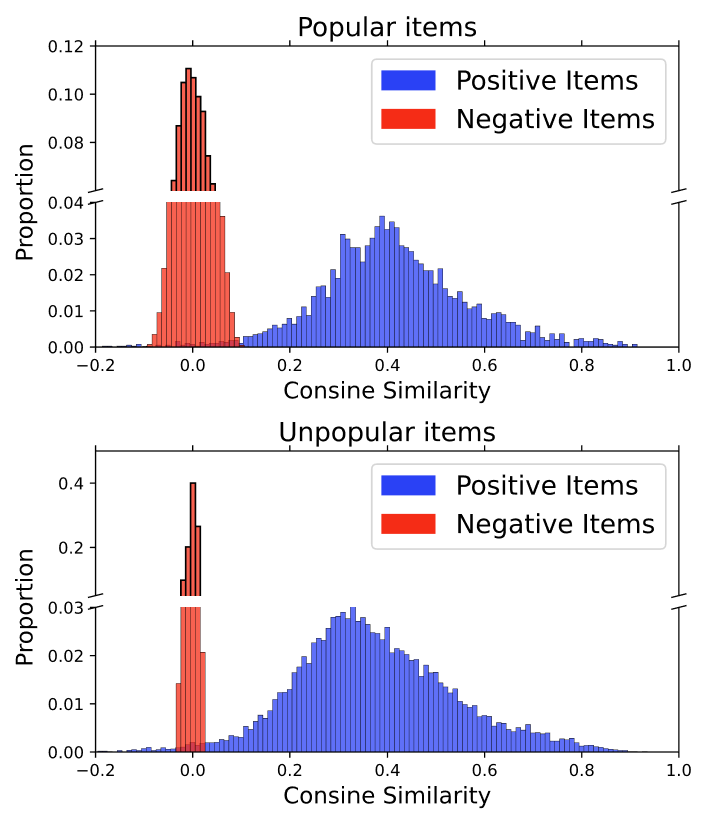}
         \caption{BPR loss (Gowalla)}
         \label{fig:BPR_dist}
     \end{subfigure}
     \begin{subfigure}[b]{0.23\textwidth}
         \centering
         \includegraphics[width=\textwidth]{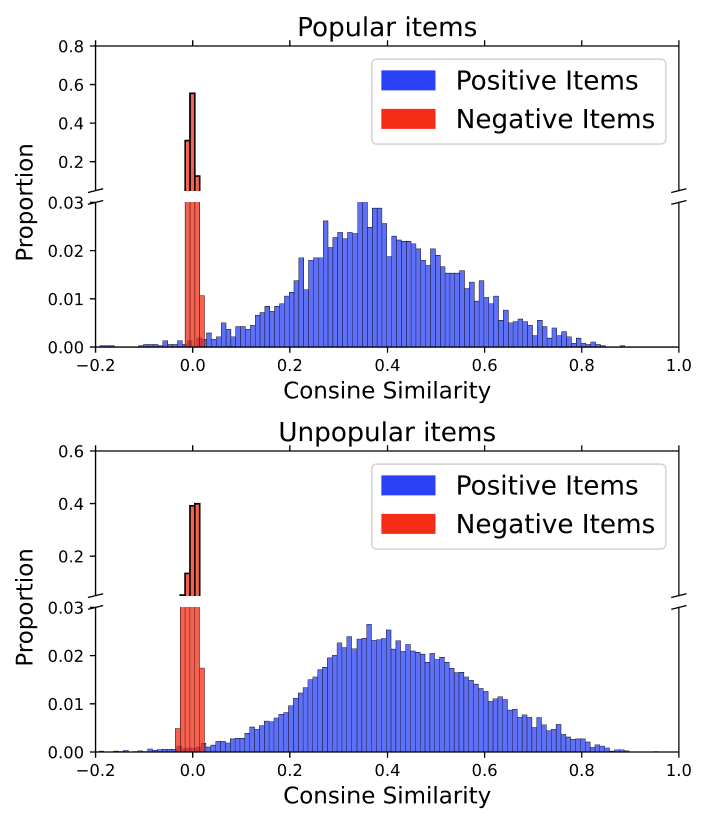}
         \caption{SSM loss (Gowalla)}
         \label{fig:SSM_dist}
     \end{subfigure}
     \begin{subfigure}[b]{0.23\textwidth}
         \centering
         \includegraphics[width=\textwidth]{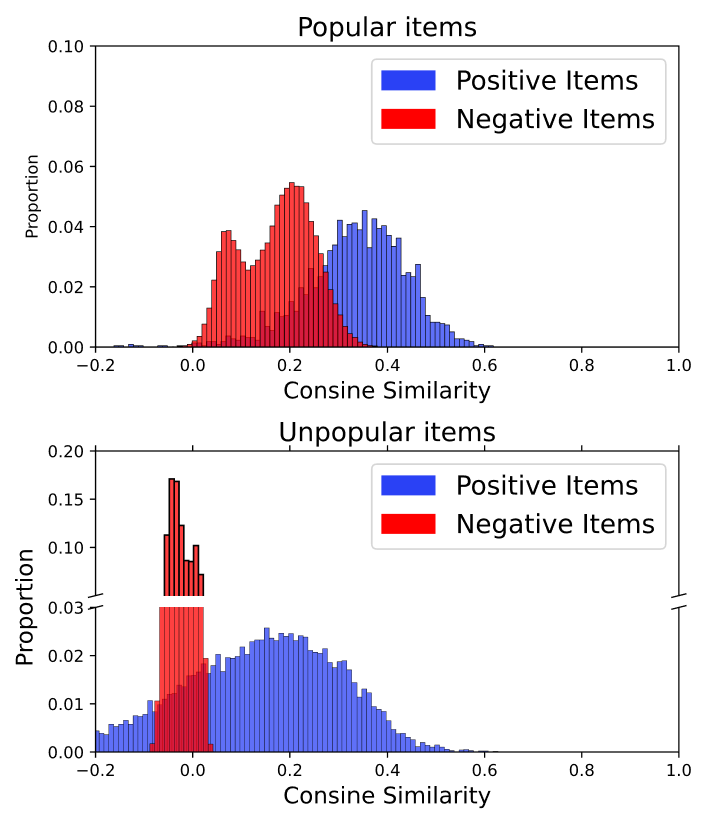}
         \caption{BPR loss (ML10M)}
         \label{fig:ML_BPR_dist}
     \end{subfigure}
     \begin{subfigure}[b]{0.23\textwidth}
         \centering
         \includegraphics[width=\textwidth]{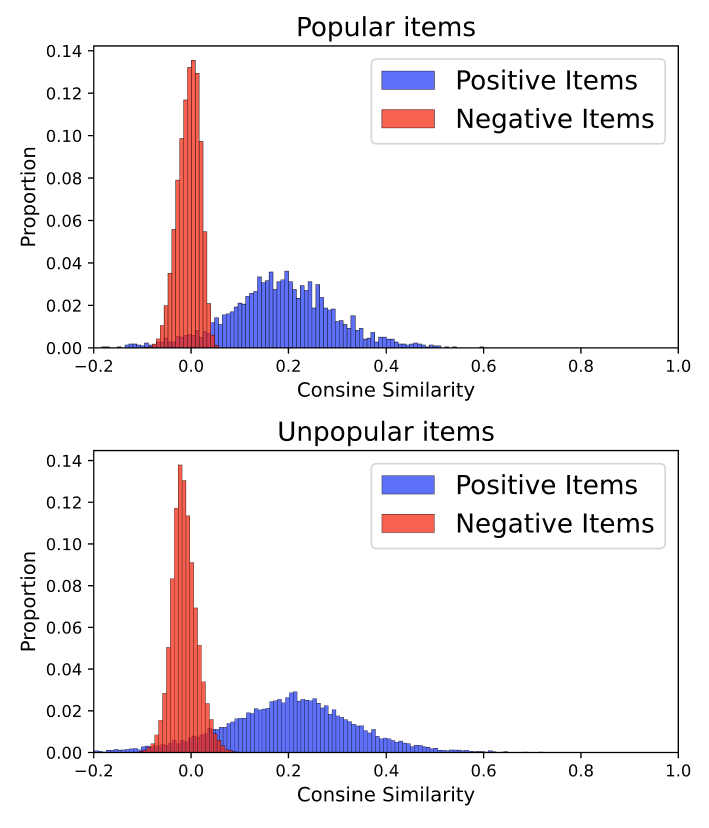}
         \caption{SSM loss (ML10M)}
         \label{fig:ML_SSM_dist}
     \end{subfigure}
        \caption{Distributions of average cosine similarity between a user and items broken down by item popularity and preference. The top figure depicts the results of the popular item group, and the bottom depicts the result of the unpopular item group.}
        \label{fig:gowalla_dist}
\end{figure*}

\cref{tab:main_results} provides the overview of the performance of baseline models and our proposed methods. We observe that \model{} yields better performance with the SSM loss compared to the BPR loss. Although it has been shown that SSM loss is robust to the bias mitigation in theory \citep{wu2022effectiveness}, our results reveal that the SSM loss still suffers from the popularity bias. 
\model{} with SSM demonstrates significant improvements over all baseline models. We outperform previous state-of-the-art approaches with a substantial margin of 4.26\% to 31.5\%  in terms of Recall@20. These results highlight the effectiveness of our approach in the mitigation of popularity bias. 
Given that \model{} does not have any additional modules during the training process, our approach is faster and more efficient compared to the other  methods. One relevant method to our approach is GRAD~\cite{ren2022mitigating}, which involves intervention in the inference stage by utilizing the accumulated gradient. However, our results demonstrate that utilizing the magnitude of the item embedding is sufficient and more efficient than utilizing the gradient of an item. 

\cref{fig:item_group_freq} shows the frequency and recall of recommended items in each popularity group. 
We divide the items into five groups according to their popularity, from the most popular (5) to the least popular (1). All item groups are arranged to have the same number of items.
Our method exhibits a superior ability to achieve fairness in recommendation results compared to LGN, which produces unnormalized output. The recommended frequency of each item group is approximately the same with \model, whereas the recommendation list of the unnormalized output is highly biased towards the popular group. Furthermore, \model{} surpasses the unnormalized output in terms of Recall@20 for the item groups from (1) to (4). The result shows the ability of our model to effectively recommend less popular items, mitigating the adverse effects of popularity bias.

\subsection{Anaylsis of the Relationship between Popularity and Embeddings (RQ2)} In this section, we analyze the relationship between embeddings and popularity and the success behind our method. 

\textit{Relationship between Popularity and Magnitude.} \cref{tab:pearson correlation} shows the pearson correlation between popularity and magnitude of item embedding. Regardless of the dataset and loss type, the magnitude of item embedding and popularity show a positive correlation.
\begin{table}[t!]
    \caption{Pearson correlation between item embedding magnitude and popularity.}        
    \label{tab:pearson correlation}
    \centering
    \begin{tabular}{c c c c c}
    \toprule
    \text { Dataset } & \text { Gowalla } & \text { Yelp2018 } & \text { ML10M }\\
    \midrule
    \text { BPR } & 0.6216 & 0.6945 & 0.7060 \\
    \text { SSM } & 0.4701 & 0.7303 & 0.8358 \\
    \bottomrule
    \end{tabular}
    \vspace{-1mm}
\end{table}

\textit{Relationship between Popularity and Cosine Similarity.} 
As \model{} leverages cosine similarity in the recommendation process, we investigate whether the cosine similarity differentiates the popularity and preference between the item and the user embedding. For the experiment, we categorize items into four different groups according to their popularity and preference for each user. We assign the most popular 20\% items into the popular group and the remaining items into the unpopular group. For preference, we divide items into positive and negative groups based on whether the item is included in the test set. As a result, we obtain the following four groups: positive popular, negative popular, positive unpopular, and negative unpopular. Then, we measure the average cosine similarity between each group and the user. 

The distribution of the average cosine similarity in each group is shown in \cref{fig:gowalla_dist}. We observe that cosine similarity effectively distinguishes the positive and negative items regardless of popularity in most cases. 
These results provide an explanation for the effectiveness of embedding normalization, particularly in the context of the unbiased dataset with the SSM loss. With BPR loss, popular item group shows high cosine similarity even if it is irrelevant to the user, suggesting the application of \model{} to SSM loss.

\definecolor{color1}{HTML}{D9A491}
\definecolor{color2}{HTML}{FF8080}
\definecolor{color3}{HTML}{FF2E4B}
\definecolor{color4}{HTML}{CC2929}
\definecolor{color5}{HTML}{801111}

\begin{figure}[t!]
    \centering
    \begin{subfigure}{.475\linewidth}
    \centering
\begin{tikzpicture}[scale=0.9, transform shape]

\begin{axis}[
    ticklabel style = {font = {\fontsize{6 pt}{6 pt}\selectfont}},
    yticklabel style={
        /pgf/number format/fixed,
        /pgf/number format/precision=5
    },
    label style = {font = {\fontsize{8 pt}{12 pt}\selectfont}},
    legend style = {font = {\fontsize{6 pt}{12 pt}\selectfont}, nodes={scale=0.7, transform shape}},
    xlabel={$p$},
    ylabel={Recall@20},
    ymin=0, ymax=0.12,
    xtick={0,0.2,0.4,0.6,0.8,1.0},
    ytick={0, 0.03, 0.06, 0.09, 0.12},
    legend pos=north east,
    ymajorgrids=true,
    grid style=dashed,
    width = 4.5cm,
    height = 4cm,
    anchor=north east, 
    line width=0.5pt,
    mark size=1pt,
    draw=none,
    fill=none,
]

\addplot[
    color=color1,
    mark=star,
    ]
    coordinates {
    (0.0,0.0222)(0.1,0.0254)(0.2,0.0303)(0.3,0.0346)(0.4,0.0398)(0.5,0.0457)(0.6,0.0527)(0.7,0.0606)(0.8,0.0692)(0.9,0.0771)(1.0,0.0853)
    };
\addplot[
    color=color2,
    mark=square,
    ]
    coordinates {
    (0.0,0.0261)(0.1,0.031)(0.2,0.036)(0.3,0.0413)(0.4,0.0473)(0.5,0.0545)(0.6,0.0619)(0.7,0.0703)(0.8,0.0763)(0.9,0.0837)(1.0,0.0906)
    };
\addplot[
    color=color3,
    mark=*,
    ]
    coordinates {
    (0.0,0.0414)(0.1,0.0456)(0.2,0.0509)(0.3,0.0572)(0.4,0.0629)(0.5,0.0672)(0.6,0.0729)(0.7,0.0782)(0.8,0.0826)(0.9,0.0838)(1.0,0.0853)
    };
\addplot[
    color=color4,
    mark=+,
    ]
    coordinates {
    (0.0,0.0621)(0.1,0.0668)(0.2,0.0713)(0.3,0.0766)(0.4,0.0822)(0.5,0.0862)(0.6,0.0877)(0.7,0.0891)(0.8,0.0884)(0.9,0.086)(1.0,0.0808)
    };
\addplot[
    color=color5,
    mark=o,
    ]
    coordinates {
    (0.0,0.1057)(0.1,0.1063)(0.2,0.105)(0.3,0.1037)(0.4,0.1)(0.5,0.0942)(0.6,0.086)(0.7,0.0758)(0.8,0.0656)(0.9,0.0526)(1.0,0.0401)
    };

\end{axis}
\end{tikzpicture}
 \caption{Recall@20}
\end{subfigure}
\begin{subfigure}{.475\linewidth}
    \begin{tikzpicture}[scale=0.9, transform shape]
\begin{axis}[
    ticklabel style = {font = {\fontsize{6 pt}{6 pt}\selectfont}},
    label style = {font = {\fontsize{8 pt}{12 pt}\selectfont}},
    legend style = {nodes={scale=0.7, transform shape}, font = {\fontsize{6 pt}{12 pt}\selectfont}},
    xlabel={$p$},
    ylabel={Frequency Ratio},
    ymin=0, ymax=1.0,
    xtick={0,0.2,0.4,0.6,0.8,1.0},
    ytick={0,0.2, 0.4, 0.6, 0.8},
    legend pos=north east,
    ymajorgrids=true,
    grid style=dashed,
    width = 4.5cm,
    height = 4cm,
    anchor=north east, 
    line width=0.5pt,
    mark size=1pt,
    draw=none,
    fill=none,
]

\addplot[
    color=color1,
    mark=star,
    ]
    coordinates {
    (0,0.0192)(0.1,0.0232)(0.2,0.0288)(0.3,0.0357)(0.4,0.0443)(0.5,0.0552)(0.6,0.0693)(0.7,0.0863)(0.8,0.1064)(0.9,0.1291)(1.0,0.1539)
    };
\addplot[
    color=color2,
    mark=square,
    ]
    coordinates {
    (0.0,0.0312)(0.1,0.0373)(0.2,0.0446)(0.3,0.0537)(0.4,0.0647)(0.5,0.0783)(0.6,0.0940)(0.7,0.1123)(0.8,0.1322)(0.9,0.1536)(1.0,0.1748)
    };
\addplot[
    color=color3,
    mark=*,
    ]
    coordinates {
    (0.0,0.0569)(0.1,0.0656)(0.2,0.0759)(0.3,0.0881)(0.4,0.1020)(0.5,0.1179)(0.6,0.1349)(0.7,0.1532)(0.8,0.1708)(0.9,0.1867)(1.0,0.2007)
    };
\addplot[
    color=color4,
    mark=+,
    ]
    coordinates {
    (0.0,0.1239)(0.1,0.1366)(0.2,0.1503)(0.3,0.1652)(0.4,0.1809)(0.5,0.1966)(0.6,0.2113)(0.7,0.2226)(0.8,0.2308)(0.9,0.2338)(1.0,0.2313)
    };
\addplot[
    color=color5,
    mark=o,
    ]
    coordinates {
    (0.0,0.7686)(0.1,0.7373)(0.2,0.7004)(0.3,0.6573)(0.4,0.6081)(0.5,0.5520)(0.6,0.4906)(0.7,0.4256)(0.8,0.3599)(0.9,0.2968)(1.0,0.2394)
    };
\legend{Group1, Group2, Group3, Group4, Group5}
\end{axis}
\end{tikzpicture}
 \caption{Frequency}
\end{subfigure}

\caption{Changes of recall@20 and the frequency ratio of each item groups with varying normalization strength $p$.}
\label{fig:norm_scale}
\vspace{-5mm}
\end{figure}
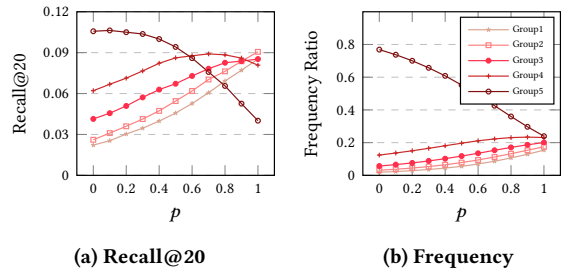

\subsection{Analysis on Scale of Normalization (RQ3)}
We conduct experiments with a varying value of $p$ that controls the normalization strength as shown in \cref{eqn:p_scale}. \cref{fig:norm_scale} shows the recall and recommendation frequency of each item group. We observe that as the normalization strength increases, the recall and frequency of items in unpopular groups increase. Note that popular groups become less dominant, and unpopular groups become dominant as we increase the strength, showing the trade-off between popularity groups. 
Our results demonstrate the potential of our method to flexibly control the impact of popularity during the inference stage. This can be further used in the real-world setting, such as tailoring the user experience to account for varying levels of popularity bias.

\section{Conclusion}\label{conclusion}
In this paper, we proposed test time embedding normalization to mitigate popularity bias in recommender systems. Our approach effectively addresses popularity bias by removing the effect of item embedding magnitude, which is highly correlated with popularity. Through extensive experiments, we have investigated the effectiveness of the proposed method and understand the impact of normalization on model performance and fairness.

\section*{Acknowledgement}

This work was partly supported by Institute of Information \& communications Technology Planning \& Evaluation (IITP) grant funded by the Korea government (MSIT) (No.2019-0-01906, Artificial Intelligence Graduate School Program(POSTECH)) and  National  Research  Foundation  of  Korea(NRF) grant funded by the Korea government(MSIT) (No. RS-2023-00217286)

\bibliography{ref}

\end{document}